\documentclass[varenna]{cimento}

\usepackage{graphicx}  
\usepackage{amssymb,amsfonts,amsmath}

\title{Colloidal Inclusions in Liquid Crystals}
\author{Randall D. Kamien}
\institute{Department of Physics and Astronomy\\ University of Pennsylvania \\ Philadelphia, PA 19104 \\ USA}
  
\begin{document}

\maketitle

\begin{abstract}
We outline the necessary background to understand the interaction of colloids in liquid crystals through boundary conditions and topology.  
\end{abstract}

\section{Introduction and Summary}
Liquid crystals and colloids -- two workhorses of modern soft matter are excellent bedfellows.  Indeed, the first theory of nematic ordering came from Onsager's classic paper, ``The effects of shape on the interaction of colloidal particles'' \cite{Onsager} that forever wed colloids and liquid crystals.  However, these lectures are not about colloidal liquid crystals but, rather, on the effect colloids have on liquid crystals through their surfaces.  Perhaps a better title for these lectures would be ``Controlling the Inside from the Outside: How Boundaries Create Cues for Liquid Crystals."  With that in mind, we will not only discuss colloidal interactions with liquid crystals, but boundary conditions in general.  On the one hand, we think of boundary conditions as the last (but essential!) ingredient in the posing and solution of differential equations.  In particular, we are quite used to this idea from our repetitive and tedious and repetitive study of electrostatics.  For the most part, the differential equations that can be used to model the elasticity of liquid crystals are of the same type, and boundary conditions lead to unique solutions in the interior of the sample.  However these differential equations are highly nonlinear, have complexities involving the subtle symmetries of the liquid crystal order, and suffer from a large number of adjustable parameters (the elastic constants!).  Stepping back, however, we also know that boundary conditions imply {\sl topology}.   What can that possibly mean?  It means that we can extract numbers, usually integers, from geometrically complex and spatially varying solutions.  Solutions with different values of these numbers cannot smoothly transform from one to the other.  Fortuitously, this is just what we need in classical physical systems since Newton's laws guarantee that the time evolution is smooth (enough).   This writeup is a guide for the beginner.  Once the reader feels comfortable with this material, further reading is encouraged.  The excellent review by Mermin \cite{mermin79} is hard to beat.  More specialized and collateral material can be found in other review articles \cite{Kamien02} and \cite{Alexander12}.  Some of the original papers worth reading include those by Toulouse and Kl\'{e}man \cite{km}, Volovik and Mineev \cite{VM}, Rogula \cite{r}, and J\"anich \cite{janich87}.   
\begin{figure}[t]
\includegraphics[width=\textwidth]{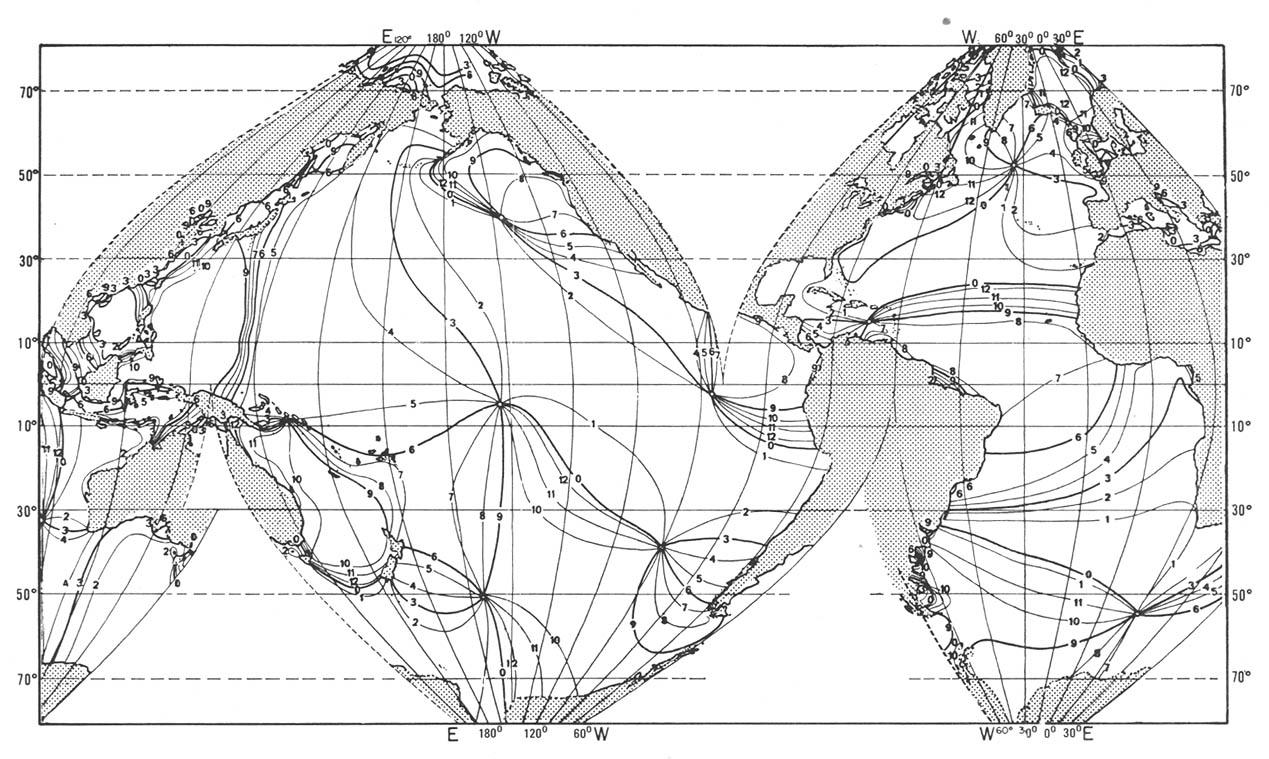}     
\caption{Cotidal lines.  Each curve goes through points that have the same high tide time.  Where they intersect it is always high tide.  From {\tt http://history.nasa.gov/SP-345/ch9.htm}.}
\end{figure}

    As a warmup, lets go back almost 200 years to the work of William Whewell, inventor of the words {\sl scientist} and {\sl physicist}.  In 1833 he used tide information to construct ``cotidal'' lines \cite{whewell}: high tide comes twice a day at a certain hour that can be measured relative to the time that the moon is directly above the prime meridian.  Assigning this time to each point on the coastline and other points where the tide is known, Whewell constructed lines of constant high tide, cotidal lines, as shown in Figure 1.  Note that on the one hand, cotidal lines cannnot cross since a point in the ocean cannot have hight tide at two different times.  On the other hand, he deduced from the known data \cite{whewell2} that there must be some point in the middle of the ocean which is {\sl always} at high tide because the cotidal lines had to all merge together. As you look at the times going around the Atlantic they go from 0 back to 12 -- where can the lines go if they don't all merge?  How can those two things be reconciled?   Recall that the tidal data does {\sl not} specify the height of the tide.  As Whewell wrote \cite{whewell2}, ``for it is clear that at a point where all the cotidal lines meet, it is high water equally at all hours, that is, the tide vanishes.''  In other words, there is a node somewhere in the ocean (also shown in Figure 1).   Conversely, if there is a point that has more than one high tide time, it must be a node and thus it is high tide there all the time and the cotidal lines all meet there.
    
 In the following lectures we will take this observation and abstract it.  We will talk about liquid crystals, materials that sometimes manifest their order by little ``clocks'' at each point in space.  We will talk about colloids (finally!) and how, when they are put in contact with liquid crystals, can impart a high tide time at each point on their boundary.  Once we get through the warmup in Flatland we will move from two to three dimensions where the nematic gets its chops.

\section{Warmup in Flatland}

So what are these liquid crystals that we have heard so much about?  On the scale of things, they are a relatively new phase of matter, discovered a little over 100 years ago (compare that to the discovery of earth, water, and air).  Here we will talk about the nematic phase.  Unlike a simple fluid, the nematic phase has an orientation everywhere in space (the clock!).  It is easy to imagine such a phase when you grab a bunch of sticks -- they don't necessarily order into a crystal like you would see pencils in a pencil box or sardines in a sardine can but, rather, align so that they are all more or less pointing in the same direction.   There are some subtleties worth mentioning but also ignoring at this point.  The nematic phase is {\sl not} polar; the direction is like pieces of chalk (and unlike sardines), no head or tail.  This is not problematic if we are studying nematics in Flatland where the direction of each chalk stick must lie in, for instance, the $xy$-plane.  In fact, most of us have dealt with this ``lack of head or tail'' our whole lives.  Look at Figure 2.  Even though the day is 24 hours long, we identify the times that are out of phase by $\pi$ -- 12:00PM and 12:00AM look the same on a civilian watch or clock.  When the hour hand goes round once we have only wound through half the day.  The civilian clock winds in half days, not for the same reason as tides (which really do occur twice a day) but because we only put 12 marks on the clock.  If we had a 24 hour clock face with a double hour hand that pointed at 0100 and 1300, 0200 and 1400, and so forth this would be the same as the nematic.  The point is that the nematic in two dimensions can wind around, just like the cotidal lines, and we can count its winding in half days, or units of $\pi$.  It can go around forward or backward but we can always keep track of it and measure the net winding as the hour hand starts and finishes at 12:00 o'clock.  The number of times its winds, with direction, is the winding number.  

\begin{figure}[t]
\includegraphics[width=\textwidth]{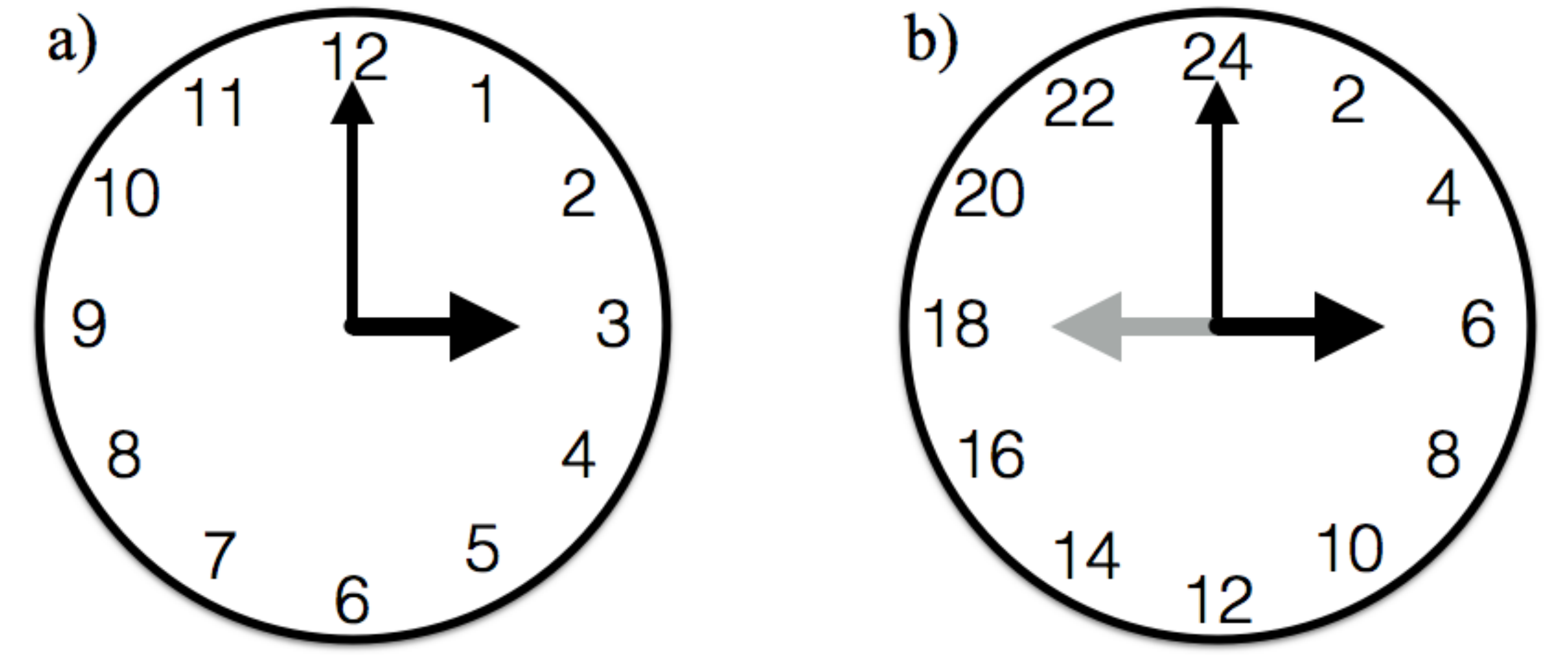}    
\caption{Clocks.  In a) we have a clock at 3 o'clock.  AM or PM?  In b) we have a military clock with all 24 hours.  Civilians identify 0600 and 1800.  If the hour hand were all one color it would be left/right symmetric and would be just like a nematic with 0600 and 1800 being equivalent.  The hour hand in a) winds around $2\pi$ when the hour hand in b) only winds around $\pi$.  The nematic can have winding in multiples of $\pi$ but can still be thought of as a clock -- the civilian clock.}
\end{figure}

In these lectures, the colloids play an essential but subsidiary role.  As with any field of interest (even American football) there is a nomenclature that we must learn to interact with our friends and colleagues.  Lets start with the colloids.  When we write ``colloid'' here, we mean a single colloidal particle, not the whole colloidal suspension.  Colloids can be metallic, plastic, or for our purposes even water droplets in the liquid crystal.  The thing that makes them interesting is how they interact with the liquid crystal through boundary conditions.  There are three types of boundary conditions that we will consider in these lectures: {\sl homeotropic}, {\sl planar degenerate}, and {\sl planar aligned}:
\begin{itemize}
\item {\bf Homeotropic} In this case the molecules are everywhere perpendicular to the surface.  In two dimensions, a disc colloid with homeotropic anchoring would produce a winding of $+2\pi$.  The sign is positive because if we go around the disc counterclockwise the director {\sl also} rotates counterclockwise.  Had the director rotated clockwise, we would call it a negative winding.  In three dimensions, a sphere with these boundary conditions would have every molecule pointing out, the so-called hedgehog.
\item {\bf Planar Degenerate} In this case, the molecules are always tangent to the surface but can point in any direction in the tangent plane.  In two dimensions this degeneracy is uninteresting -- the molecules must just wrap around the disc and will thus force a winding of $+2\pi$ again.  In three dimensions, for a spherical colloid, we would get a bunch of lines tangent to the sphere but with no special order.  It could, for instance, represent the direction of the wind on the earth (where we ignore East versus West, or North versus South, {\sl etc.}).
\item{\bf Planar Aligned} Again, the molecules are tangent to the surface and, again, in two dimensions this just gives a $+2\pi$ winding.  In three dimensions, the difference between degenerate and aligned becomes clear.  Planar aligned anchoring not only means that molecules are tangent to the surface but that, in addition, their direction on the surface is specified.  Typically, this is achieved through mechanical rubbing of a surface coating.  For instance, if we had a spherical colloid, we could achieve anchoring where the molecules always point along the local lines of latitude or longitude.  More often, planar aligned boundary conditions are on flat surfaces or along capillaries.  We will not touch too much on this situation, if at all.
\end{itemize}

In addition to the boundary conditions on the colloid, we have boundary conditions on the boundary!  Here we will only consider these three boundary conditions for the outer boundary plus one more: the uniform state, that is where the molecules all point in one direction.  This is a mixture of the above anchoring conditions since, for example, in a cylinder we would have planar aligned on the walls with the alignment along the cylinder direction while, at the same time, we would have homeotropic anchoring on the top and bottom of the cylinder.  

What can we do with all these words?  By now, in a course on fluid mechanics or electromagnetism, we might be solving equations but here we have not even written down any equations!  Consider, however, a liquid crystal cell in the shape of a disc in two dimensions with hometropic anchoring.  This can be achieved by spreading a thin film of nematic over a circular opening on a slide treated for homeotropic anchoring.   We now have a situation almost identical to the tides.  Because the molecules are perpendicular to the boundary circle, the clock goes around by 24 hours, a winding of $2\pi$.  If we try to draw ``cotidal'' lines emanating from the boundary then we will have the same problem that Whewell had, there must be places where the cotidal lines meet.  But now things are different.  The nematic is a rod of fixed length, we don't get to talk about a place where it is always high tide.  We now have a problem -- how can the molecule at the intersection point in more than one direction (using the ambiguity of the civilian clock does not help for {\sl all} the angles)?  It can't.  We can fix the problem two ways depending on whether we love free energies or we love topology.  

\begin{figure}[t]
\includegraphics[width=\textwidth]{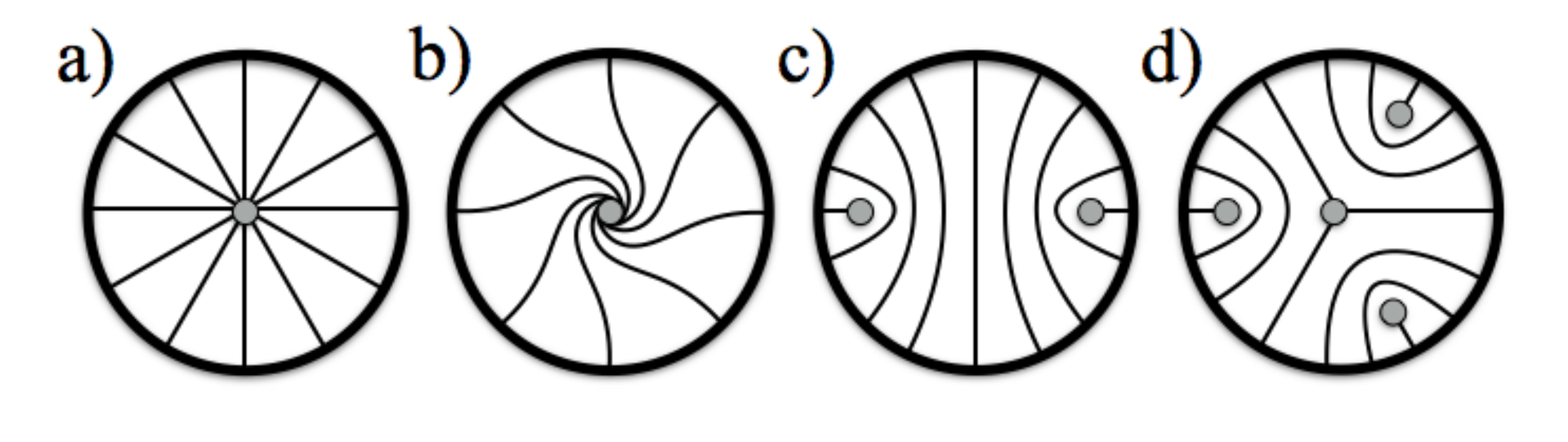}     
\caption{The nematic in the homeotropic disc.  In a) we have a configuration with homeotropic boundary conditions on the inner boundary allowing for a colloid with homeotropic anchoring.  The lines do not have to be straight: in b) we show a defect that has homeotropic anchoring on the outer boundary and planar anchoring on the inner boundary.  Both wind by $2\pi$.  In c) we get a texture that arrows or magnets cannot have -- there are defects that only wind by $\pi$.  The two defects add together to make a total winding of $2\pi$ on the outer wall.  In d) we have four defects: three that wind $\pi$ and one that winds $-\pi$ (which one and why?).  These add up to a total winding of $2\pi$.}
\end{figure}

Lets start with the topological point of view -- since it is impossible to have something point in multiple directions at once we are forced to conclude that there is some terrible discontinuity at this ``defect.''  We can fix this by excluding this point from the disc.  If $D^2$ is the two-dimensional disc and the the defect is at a point $p$, we denote the leftover space as $M=D^2\setminus \{p\}$, $D^2$ minus the set $\{p\}$.  Then, on $M$ everything is ok.  The one place where we might have had a problem is missing and the map from $M$ to the nematic clock is everywhere smooth.  We could say that we have smooth maps from the punctured disc to the clock or, stepping back, we can imagine expanding the puncture which is just a smooth distortion of the sample into an annulus and, in fact, even in to a circle.  In doing so we are forced to conclude that the winding on the inner boundary around the defect must be exactly the same as the winding on the outer boundary.  The outer boundary specifies the topology of the inner boundary.  Is this unique?  NO!  There are other ways to satisfy the outer boundary conditions. We can do something in the nematic that we couldn't do with vectors at each point; we can have half charge defects!  

Some of these possibilities are shown in Figure 3.  All four configurations have homeotropic anchoring on the outer boundary -- the direction of the molecules, depicted by the light lines, always comes in perpendicular to the circular wall.  In Fig. 3a we see the lines coming straight in, ending on a hole that could be a colloid {\sl also} with homeotropic anchoring.  The winding on the inside is $2\pi$ and the nematic lines are also cotidal lines.  The nematic clock points the same way along each radial line.  However, consider Fig. 3b: there the homeotropic anchoring on the outer wall changes to planar anchoring on the inner colloid.  The winding is {\sl still} $2\pi$ but the geometry is different.  More importantly, these lines are no longer cotidal lines as the direction of the nematic clock changes along them.  Try sketching the lines of constant nematic direction for Fig. 3b and convince yourself that topology is more robust, while geometry is more refined.   

Things get more complicated still because the nematic can have defects that wind by $\pi$.  Consider Fig. 3c: here there are two defects.  If you imagine the two defects being drawn together at the center of the disc you would wind up with the configuration in Fig. 3a.  Note, however, that the lines in Fig. 3c could {\sl never} be cotidal lines.  Why not?  Look at the line on the right ending on a colloid. Just above it the tide time is earlier and just below it the tide is later (or {\sl vice versa} since all that matters here is the difference between ``before and after'' or ``after and before.'').  As we move along that line, towards the colloid, we always have before and after perfectly well defined.  But what happens when we go around the defect on its left?  Now before and after are connected!  Unless there are no tides whatsoever, this cannot happen and, because we specify the tidal times on the outer boundary, we know there are tides.  But a nematic can do this because 1200 and 2400 are the same.  Going clockwise from 2400 hours (at the top) we first hit 0100 hours.  Going counter clockwise from the 2400 on bottom ({\sl i.e.} 1200 hours) gives 2300.  Continuing in this fashion we get to the line segment on the right.  It is both 1800 and 0600 -- impossible for cotidal lines but completely correct for nematics!  Similarly, in Fig. 3d we have four defects each of which wind by $\pi$, though the one in the center winds in the opposite direction.  Check for yourself.  
\begin{figure}[t]
\centerline{\includegraphics[width=0.9\textwidth]{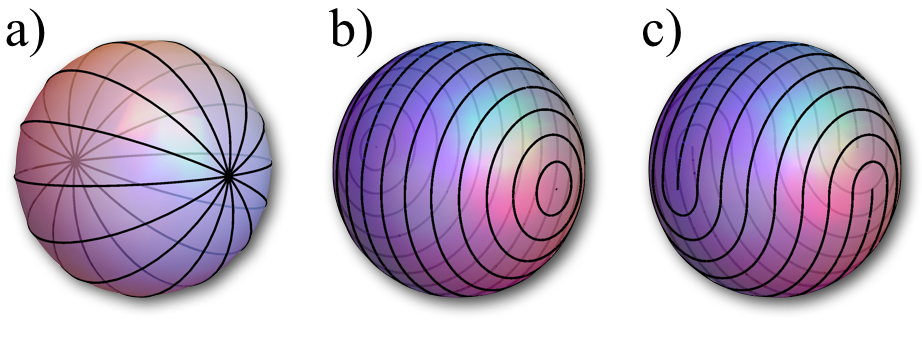}}  
\caption{Three different nematic textures on a sphere: a) lines of longitude, b) lines of latitude, c) lines of a baseball stitch.  Check that the winding of the defects in a) and b) are $2\pi$ and only $\pi$ in c). Figure after \cite{mosna}, courtesy of G.P. Alexander.}
\end{figure}

The half defects, however, do not have an easy interpretation in terms of colloids with specific anchoring.  Indeed the half-winding defects do not have homeotropic or planar anchoring and cannot for, if they did, they would be integer-winding defects.  So are these forbidden?  No, this is where we come to the other way of dealing with defects, namely through the love of free energies.  Recall that the nematic director (the direction on the clock) never vanishes so we cannot have all the lines meet.  When there are colloids in the sample, we can deal with this since there is no nematic field on the inside of the colloids and we have poked a hole.  But can we revive the argument about cotidal lines?  Yes.  The nematic is only one phase of the liquid crystal material.  At higher temperatures, for instance, the same molecular fluid becomes isotropic with the nematic director completely random.  As is the usual procedure, we can quantify the amount of nematic order through an order parameter, in this case the {\sl Maier-Saupe order parameter} denoted by $S$ \cite{maier}.  If the director field points on average in the some direction $\hat{\bf n}_0$ then $S\equiv {1\over 2}\langle\,3\cos^2\theta -1\,\rangle$ where $\theta$ is the angle between the actual director and the average director and $\langle\cdot\rangle$ is the thermodynamic average.  Note that if the molecules are disordered so that $\theta$ is distributed uniformly between $0$ and $\pi$, then $\int_0^\pi \sin\theta d\theta\,\cos^2\theta = {1/3}$ and $S=0$ -- no order.  If all the molecules point along $\hat{\bf n}_0$ then $S=1$.  One can, using the procedures of Landau theory, construct a free energy that depends on $S$ (and it would give a first order phase transition).  The point is, that by allowing $S$ to vary in space along with the director $\hat{\bf n}$ we can revive the cotidal lines -- we can let $S$ vanish where the defect would be so that there is no nematic order there.  In effect, we have poked a hole in the nematic.  Thus we can have half-winding defects if we just let the order parameter vanish at those points.  Those defects we might expect to be much smaller since the scale of the hole is set by molecular parameters and lengthscales.  More exotic colloids can also sit in those locations if they have mixed anchoring or no anchoring.

So ends our discussion in two dimensions.

\section{The Boojums}
\begin{figure}[t]
\centerline{\includegraphics[width=0.5\textwidth]{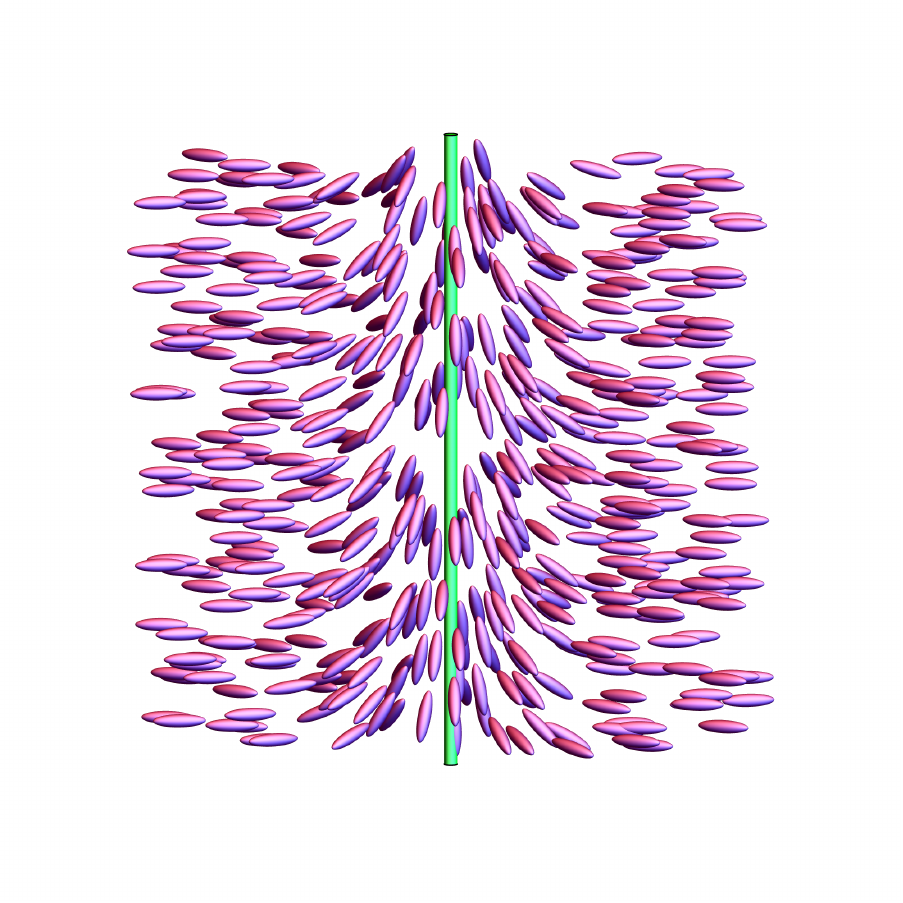}}  
\caption{Escape of a $+2\pi$ winding into the third dimension.  Note that away from the center of the figure, the molecules take on the boundary conditions in Fig. 3 but, by the time they are at the center, have found a common way to point. Figure courtesy of D.A. Beller.}
\end{figure}
As we move up in dimension, visualization becomes more difficult and so a little contortion and gymnastics will be used.  First we will consider planar degenerate boundary conditions on a sphere and then homeotropic anchoring on a sphere.  Alone, this colloidal boundary condition is very simple but, in concert with the surrounding medium, we will see that this leads to some interesting structures.  

Planar degenerate anchoring means that the nematic director is everywhere tangent to the surface of a colloidal sphere.  It could, for instance, represent the wind direction on the surface of the Earth.  There is a problem here: suppose that the wind were blowing East at every point; which way does it point at the North and South poles?  Suppose the wind were blowing North at every point; which way does it point at the poles again?  It turns out that if you try to put a two-dimensional vector at every point on a sphere then there must always be {\sl at least} two points that have winding $2\pi$.  If it is a nematic on a sphere then you can have four points of $\pi$ winding.  Just like in the case of Figure 3d, however, we can have extra defects so long as they all add up to $4\pi$.  This is a very deep result from topology that plays an overarching and essential role in the study of inclusions in nematics.  This can be proved using calculus ({\sl lots} of calculus) but we will not do it here.  See \cite{Kamien02} for a physicist's approach to the proof.  For our purposes then, we can think of a nematic field tangent to the sphere making a texture like the lines of longitude as in Fig. 4a, lines of latitude as in Fig. 4b, or perhaps like the set of lines in Fig. 4c where there are {\sl four} half-winding defects.  

\begin{figure}[t]
\includegraphics[width=\textwidth]{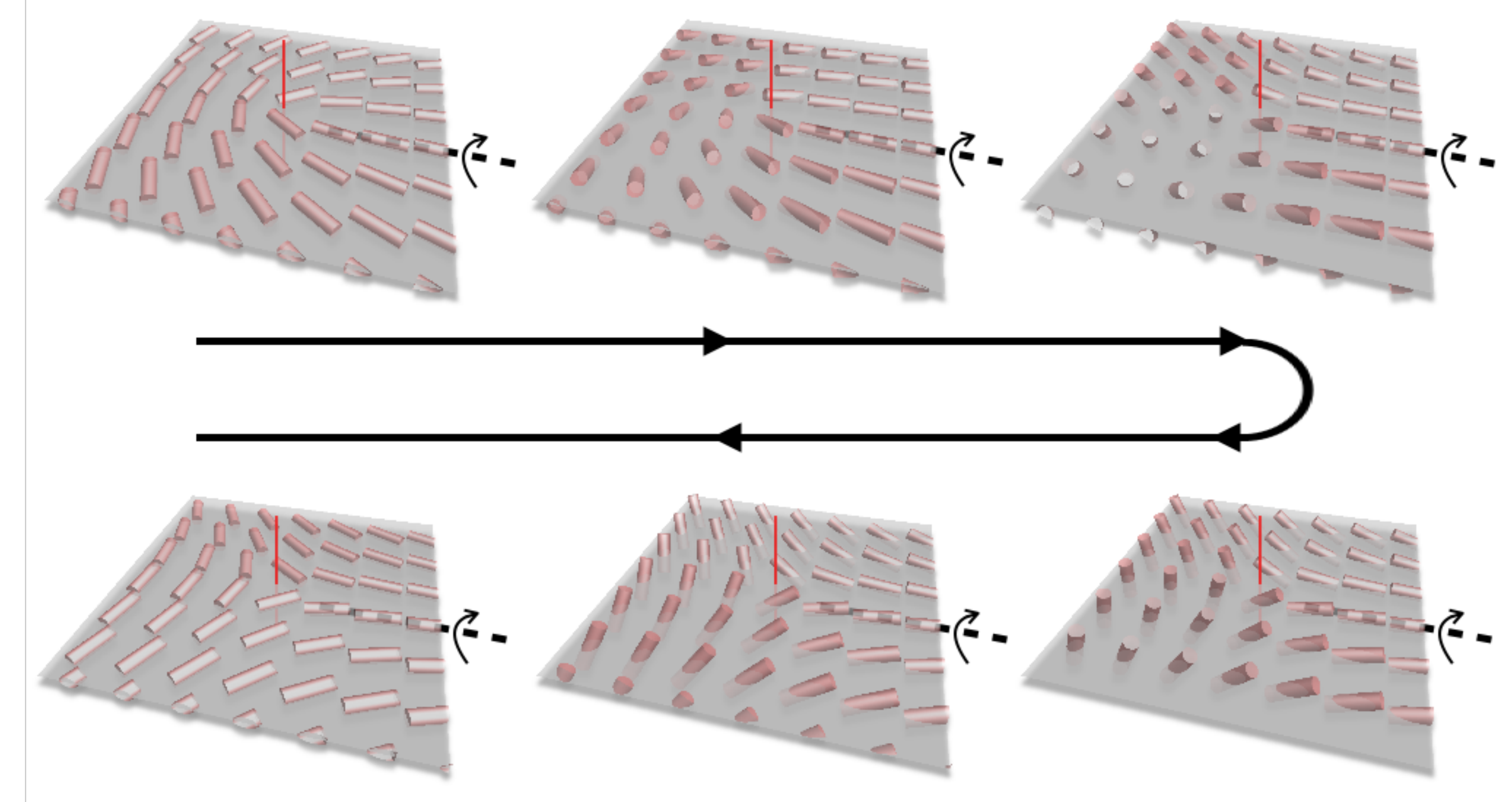}     
\caption{The complication of nematics: Going clockwise from the upper left we see that we can rotate every cylinder about the denoted axis by $\pi$ and continuously change something with winding $+\pi$ to winding $-\pi$.  Note that we are rotating each cylinder around its own center and not in space -- each cylinder, however, is rotated around the same axis fixed in space.  The cylinder that is pierced by the dashed line (the axis) rotates but is symmetric about that line so nothing happens.  The cylinder in the middle of the left side of the sample rotates by $\pi$ around a short axis so that in the fourth frame it is actually pointing up. The result of this smooth transformation implies that these two textures (and all the textures in between) are topologically equivalent.  Figures courtesy of G.P. Alexander.}
\end{figure}

If we were just interested in studying the nematic on the surface,we would be done.  However, this colloid is in a three-dimensional nematic ocean.  How do these defects on the surface exert their topology into the bulk (remember -- ``controlling the inside from the outside'')?  In the case of the $2\pi$ winding, for instance off the North pole of the lines of longitude texture in Fig. 4a, one might expect some sort of vortex line with the same winding emanating from the surface.  While that may be the case close to the surface, the nematic director, now able to explore all three dimensions, can escape \cite{meyer}!  As shown in Figure 5, while the director may go round by $2\pi$ away from the center of the defect, as in Fig. 3a, it can start pointing up so that by the time we get to the defect core at the center, there is no defect!  The molecules know exactly which way to point and they point up.  The $2\pi$ winding has disappeared into the bulk leaving nothing behind.  It turns out, however, that the half-winding defects survive into the bulk.  In Figure 6, we show the escape into the third dimension of a $\pi$ winding defect to a $-\pi$ winding defect.  This is a continuous deformation and so these two geometries are topologically the same.  In fell swoop, however, this picture explains everything about the defect lines in bulk.

In the case of half-winding defects, we cannot massage them away like we can in Figure 5.  When we try to, we discover that we just get another half-winding defect, apparently of the opposite sign!  How can $+\pi$ turn to $-\pi$ continuously?  It cannot -- it was wrong to assign a sign to the defects.  There is either a half-winding defect or there is not a half-winding defect.  What about the integer-winding $2\pi$ defect?  Well, in a planar cut we might start with a texture as in Fig. 3a.  However, the $2\pi$ winding can be divided up into two $\pi$ windings as in Fig. 3c.  Since we are allowed now to rotate the director out of the plane, we can smoothly change one of the two defects to look like a $-\pi$ winding and thus the net winding in the two-dimensional slice would appear to be $0$.  A fancy way of saying this is ``the fundamental group of the three-dimensional nematic is the group $\{+1,-1\}$ under multiplication.''  That is, the things we can measure by calculating the winding around a circle (the fundamental group) are either nothing, $+1$, or something, $-1$.  Two somethings gives $(-1)^2=1$ -- nothing!  If you are short for space, you can write $\pi_1(\mathbb{R}P^2)=\mathbb{Z}_2$ \cite{mermin79}, but I have no page limit.
\begin{figure}[t]
\centerline{\includegraphics[width=0.7\textwidth]{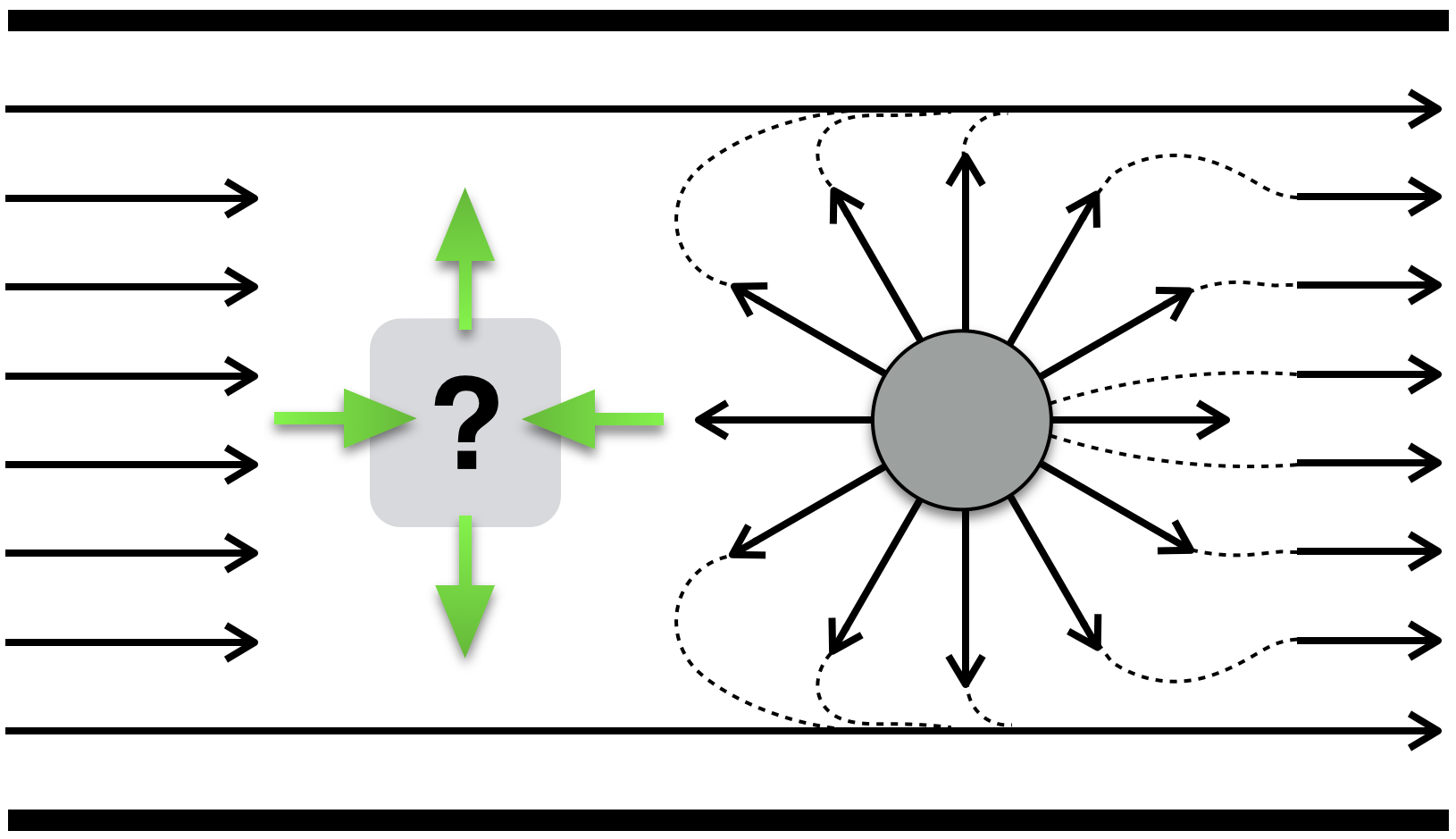}  }   
\caption{The mystery of the hedgehog in a uniform field.  The solid lines give us the boundary conditions while the dashed lines try to fill in the space.  In this situation we forbid any half-winding defects.  There must be a zero of the director field in the box, so there must be a defect inside! }
\end{figure}
What does this all mean?  Colloids with planar anchoring only interact with the bulk when they have half-winding defects on them.  The texture in Fig. 4c would have to have four threads of the half-winding defect attached to the surface defects.  It has been proposed that this could lead to the assembly of a tetravalent colloidal chemistry \cite{nelsonnano}.  The threads can either connect one colloid to another or they can even connect to themselves making two handles in the bulk, for instance.  If someone gets fancy and chooses planar aligned anchoring on the spherical colloid, then they must place the defects too, allowing them to control the shape of the ``atomic valence.''
One place where planar degenerate anchoring is used is in capillaries.  Often, for energetic reasons, the nematic director points along the long axis of the tube.  We will need this boundary condition to set up the far field in the next section.

\begin{figure}[t]
\centerline{\includegraphics[width=0.7\textwidth]{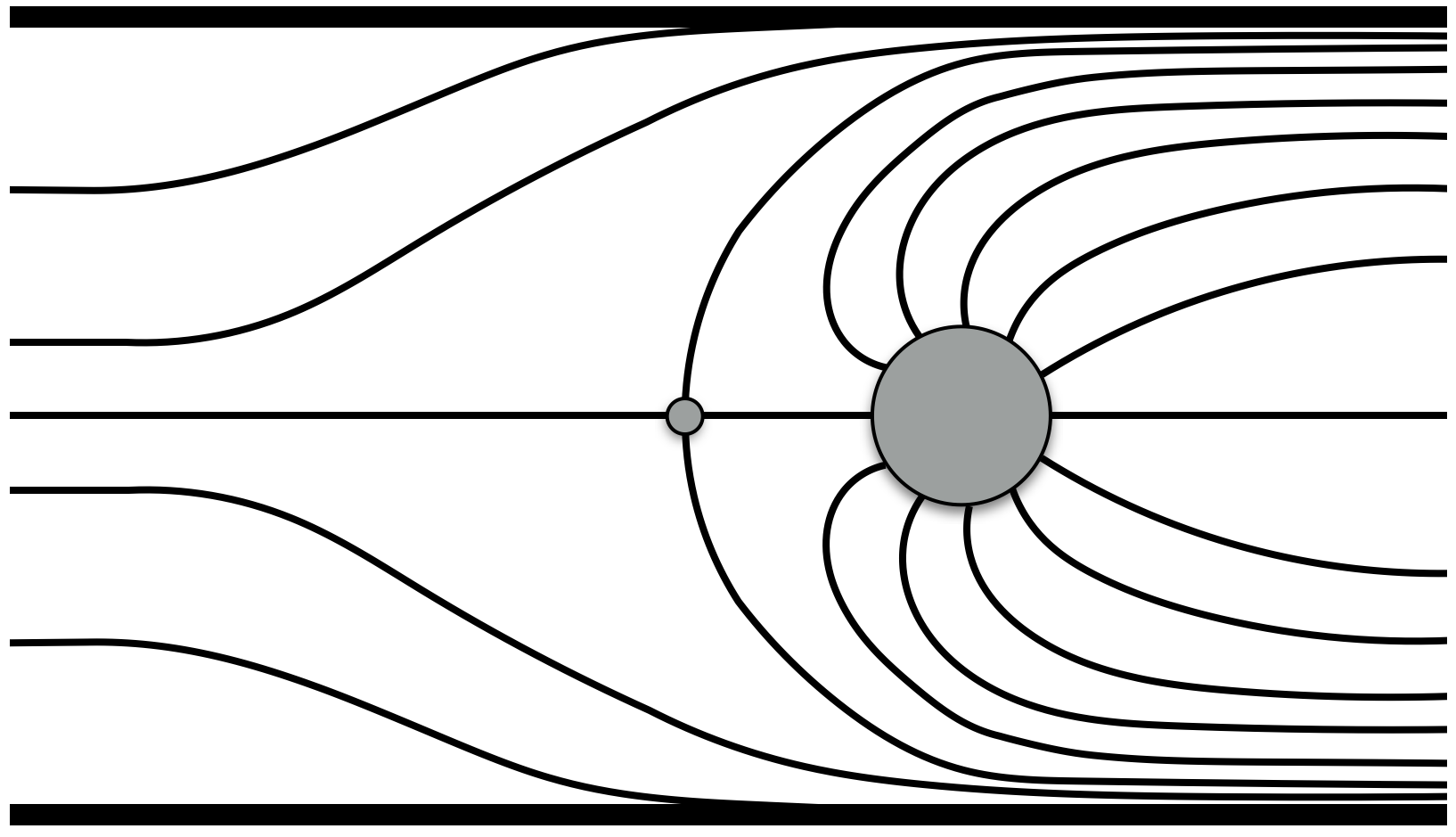}   }  
\caption{Mystery solved.  The big colloid forces a small point defect where the order parameter must vanish.}
\end{figure}

\section{The Hedgehog}

Now we consider the situation where a spherical colloid has homeotropic anchoring so that the director points out in every direction, like the spines of a hedgehog.  We could always put it in a larger spherical container with the same boundary conditions and end up with a three-dimensional version of Fig. 3a.  However, it is more interesting (and easier to set up experimentally) to consider a hedgehog in a uniform background.  We can achieve this by having a very long capillary with planar degenerate anchoring along the tube or we could have very large plates above and below the colloid with homeotropic anchoring.  It is sometimes useful to think of these boundary conditions in terms of electric field lines \cite{poulin97}, but it is also easy to get fooled by the analogy.  We will have to do our best.  Suppose we have the situation shown in Figure 7 and we first consider a situation where there are no half-winding defects.  In that case, we can make a global choice of direction and never need worry about the director changing from up to down (no defects!).  At the left and right and on top and on bottom, the director is uniformly pointing towards the right.  At the surface of the colloid, the director points out everywhere.   Now, lets try to connect all the arrows so that we go smoothly from the surface of the sphere to the boundary at ``infinity.''  We can do it on the right without too much trouble and, as we move to the left of the colloid we can still do it by bending the director around.  But look at the mystery box?  We can see that on top the director will have an up component and on bottom it will have a down component.  Somewhere between the top and bottom there must be a line (that need not be straight) where the vertical component of the director vanishes.  Likewise, just from what we see, there must be some line (again, not necessarily straight) that separates the right pointing director on the left from the left pointing director on the right.  But these two lines must cross somewhere in the box.  ACK!  This means that the director must vanish inside the box since it will have no horizontal {\sl and} no vertical component.  We are forced to conclude that there must be a zero inside and therefore there must be a defect where the director does not know which way to point.  To calculate the charge of this new defect is beyond these lectures but the reader is encouraged to study \cite{mermin79}, \cite{Alexander12}, or \cite{poulin97}.  It turns out, not surprisingly that if we assign charge ``$+1$'' to the colloid then the new defect has charge ``$-1$'' leaving us with a neutral situation as in Figure 8.

\begin{figure}[t]
\centerline{\includegraphics[width=0.6\textwidth]{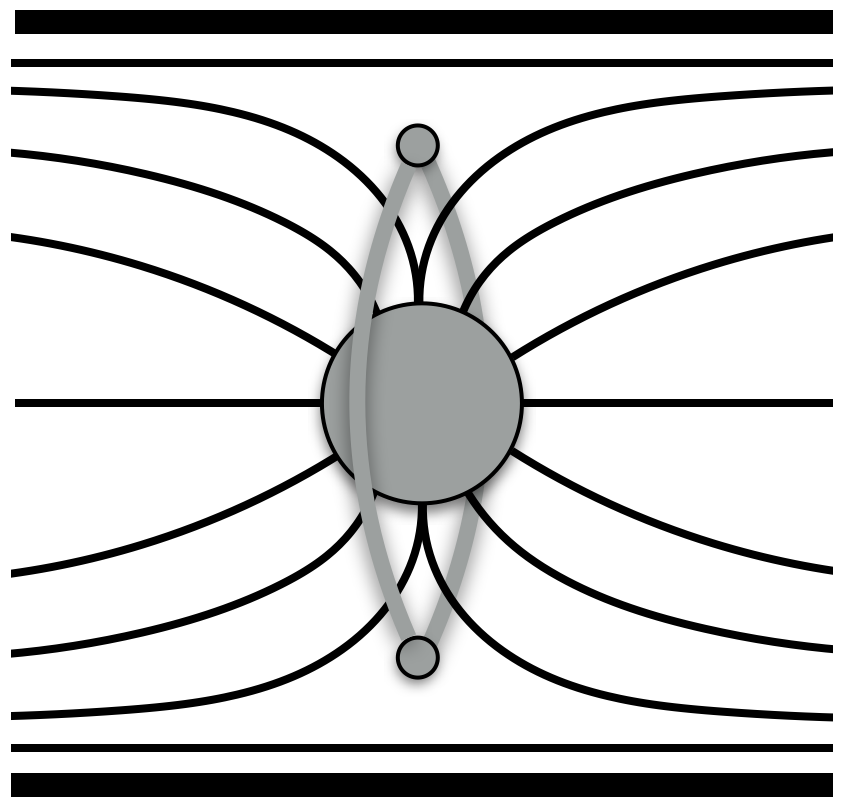}   }  
\caption{The Saturn Ring.  The colloid has hometropic anchoring and the boundary conditions at infinity are parallel to the horizontal axis.  Note that now there are two half-winding defects.  In three dimensions, these defects turn into a ring surrounding the colloid.  Note that in this configuration it is impossible to put arrows on the lines because of the $\pi$ winding around the Saturn Ring.}
\end{figure}

What if we allow for half-winding defects?  In a prescient paper by Terentjev \cite{terentjev95}, it was suggested that a ``Saturn Ring'' defect could form -- something that mixes line defects and hedgehogs.  This texture is shown in Figure 9.  In this case a loop of $\pi$ winding encircles the colloid.  The geometry of the nematic texture still allows parallel boundary conditions at infinity.  What may be surprising is that the point defect in Figure 8 can be replaced by a loop of line defect in this case.  Loops can have both half-winding in a transversal plane but can {\sl also} behave like a point defect \cite{Alexander12}.  Two for one!

In recent years, the ability to control, create, and manipulate these line defects around colloids and other geometric cues has led to a rich set of phenomena ranging from the tying of knots \cite{zumer}, to the sculpting of elastic distortion fields \cite{gharbi}, to the threading of wires \cite{galerne}, to the complex wiring of networks in porous media \cite{serra}.

If Whewell only knew.

\acknowledgments
These lectures distill many years of work and collaboration with many talented individuals including Gareth P. Alexander, Daniel A. Beller, Marcello Cavallaro, Jr., Bryan G. Chen, Simon \v{C}opar, Mohamed A. Gharbi, Iris B. Liu, Yimin Luo, Thomas Machon, Elisabetta A. Matsumoto, Carl D. Modes, Ricardo A. Mosna, Christian D. Santangelo, Francesca Serra, Daniel M. Sussman, Kathleen J. Stebe,  Lisa Tran, Vincenzo Vitelli, Yu Xia, Shu Yang, and Primoz Ziherl. I am indebted to them for countless discussions, numerous arguments, and many glasses of Scotch.  Particular thanks goes to G.P. Alexander and D.A. Beller who both provided me with original figures for this manuscript and Max Lavrentovich, I.B. Liu, Y. Luo, F. Serra, and L. Tran for critical reading and comments.  This work was partially supported by NSF DMR12-62047, the University of Pennsylvania MRSEC Grant NSF DMR11-20901, and by a Simons Investigator grant from the Simons Foundation.

\end{document}